\documentclass[preprint]{revtex4}
\usepackage[T2A]{fontenc}
\usepackage[cp1251]{inputenc}
\usepackage[english]{babel}
\usepackage{amsmath}
\usepackage{amsfonts}
\usepackage{graphicx}
\usepackage{epsfig}
\usepackage{subfigure}
\usepackage{amssymb}
\usepackage[indent,small]{caption2}

\begin{document}

\title{The influence of boundary conditions on the form of the optical beam
in the array of coupled optical waveguides.}

\author{Babichenko V. S.$^1$, Gozman M. I.$^{1,2}$, Guseynov A. I.$^3$, Habarova T. V.$^1$,
Pavlov A. I.$^1$, Stavcev A. Yu.$^3$, Tsyvkunova E. A.$^3$,
Polishchuk I. Ya.$^{1,2}$.}

\affiliation{$^1$ RRC Kurchatov Institute, Kurchatov Sq., 1,
123182 Moscow, Russia}

\affiliation{$^2$ Moscow Institute of Physics and Technology,
Institutskii per., 9, 141700, Dolgoprudny, Moscow Region, Russia}

\affiliation{$^3$ National Research Nuclear University MEPhI
(Moscow Engineering Physics Institute), 115409, Kashirskoe shosse,
31, Moscow, Russia}

\begin{abstract}

We investigate the optical beam behavior in the periodical array
of the coupled optical waveguides with the monotonic change of the
refractive index in the transverse direction. We consider the
dependence of the form of the optical beam on the boundary
conditions. It is well known that if the input wave packet is wide
enough, the optical Bloch oscillations occur, while for the enough
narrow input wave packet the breathing mode is observed. We show
that if the input wave packet is neither too wide nor too narrow,
the optical beam takes a peculiar form which can be considered
neither as the Bloch oscillations nor as the breathing mode. We
qualitatively explain the transformation of this intermediate form
of the optical beam when the width of the input wave packet
changes.

\end{abstract}

\maketitle


\section{Introduction}
\label{Sec:Intro}

The optical waveguides are the inherent component of the optical
and optoelectronic devices, indispensable for the optical signals
transmission between different parts of the system. The
interaction of different waveguides is usually an undesirable
effect which distorts the transmitted signal. Thus, the isolation
of the waveguides is an important problem for the optical devices
design.

However, in some cases, the interaction of the waveguides can
cause unexpected phenomena useful in practice. Those phenomena can
occur in periodic arrays of waveguides, which are a special kind
of low-dimensional photonic crystals. The main feature of these
systems is a band structure of the optical spectrum which defines
their peculiar properties
\cite{Lourtioz,Joannopoulos,Busch,Longhi}.

\begin{figure}
\centering
\includegraphics[width=0.7\textwidth]{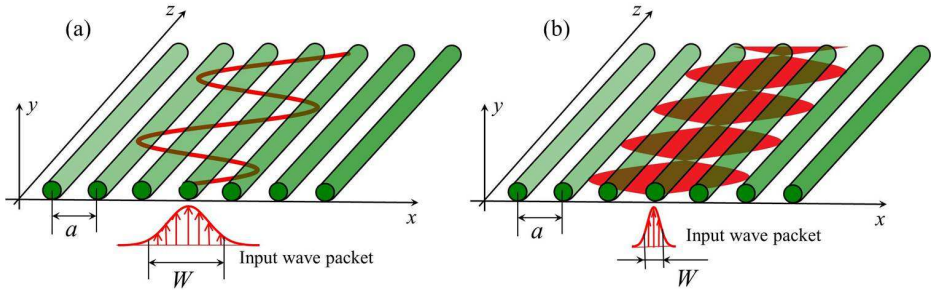} %
\caption{The array of waveguides and the optical beam:
(a) Bloch oscillations, (b) breathing mode. } %
\label{Fig01_Scheme}
\end{figure}

Among the most famous phenomena which arise due to the band
structure of the periodic arrays there are the optical Bloch
oscillations and breathing modes (see, for example,
\cite{Pertsch1998_Theor, Pertsch1999, Morandotti1999, Pertsch2002,
Chiodo2006, Gradons_Zheng2010}). These are two different forms of
the optical beam which occur in the arrays with the monotonic
change of the refractive index of the waveguides in the transverse
direction. In the case of the optical Bloch oscillations, the
width of the light beam is substantially constant, but the beam
path possesses the curved oscillatory form (see Fig.
\ref{Fig01_Scheme}(a)). In the case of the breathing mode, the
beam path is straight, but the width of the beam changes
periodically, i. e. the optical beam periodically spreads and
focuses (see Fig. \ref{Fig01_Scheme}(b)).

The form of the optical beam depends on the boundary conditions at
the edge of the array (the plane $z=0$ in Fig.
\ref{Fig01_Scheme}). The boundary conditions are specified with
the input wave packet. The optical Bloch oscillations arise if the
input wave packet possesses the Gaussian form with the width $W\gg
a$, $a$ being the period of the array. The breathing mode arises
if the input wave packet excites a single waveguide only, i.e.
$W\lesssim a$.

The purpose of this work is to find the form of the optical beam
for the intermediate case $W\gtrsim a$, when the optical beam
behavior can be considered neither as the Bloch oscillations nor
as the breathing mode. For some values of the input wave packet
width, we obtain the peculiar form of the optical beam which were
unknown before.

We apply the multiple scattering formalism for the numerical
simulation. This method was used earlier in our works
\cite{We_OptEng2014,We_ICTON2014} for the theoretical
investigation of the Bloch oscillations.

This paper is organized as follows. In Sec. \ref{Sec:MSF} we
describe the multiple scattering formalism. In Sec.
\ref{Sec:CalcMethod} we present the algorithm for calculating the
form of the optical beam in an array of waveguides for the
specified boundary conditions. In Sec.
\ref{Sec:NumericalSimulation} we apply the method explained in the
previous section for calculating the optical beam form for the
boundary conditions possessing the form of the Gaussian wave
packet. Varying the boundary conditions, we obtain the different
form of the optical beam. The obtained results are summarized in
Sec. \ref{Sec:Conclusion}.


\section{Multiple scattering formalism}
\label{Sec:MSF}

Let us consider the array of $N$ parallel infinite cylindrical
dielectric waveguides parallel to the $z$-axis. All the waveguides
are assumed to possess the same radius $R$ but different
refractive indices $n_j$, $j$ being the waveguide number. The
refractive index of the environment is $n_e$. The permeability of
the waveguide material and the environment is unity.

Suppose that a guided mode with a frequency $\omega$ is excited
within the array. Then, all the components of the electromagnetic
field are proportional to $e^{-i\omega t+i\beta z}$.

Let us consider the field of the guided mode inside of the array.
The field of the guided mode is finite in any point inside the
waveguide. So, the electromagnetic field inside of the $j$-th
waveguide may be represented in the form
\begin{equation}
\begin{array}[c]{l} \displaystyle
\tilde{\mathbf{E}}_j(\mathbf{r})= %
e^{-i\omega t+i\beta z}\,\sum\limits_{m=0,\pm1...} e^{im\phi_j}\, %
\Bigl(c_{jm}\,\tilde{\mathbf{N}}_{\omega_j\beta m}(\rho_j) %
-d_{jm}\,\tilde{\mathbf{M}}_{\omega_j\beta m}(\rho_j)\Bigr),
\medskip \\ \displaystyle
\tilde{\mathbf{H}}_j(\mathbf{r})= %
e^{-i\omega t+i\beta z}\,n_j\sum\limits_{m=0,\pm1...}e^{im\phi_j}\, %
\Bigl(c_{jm}\,\tilde{\mathbf{M}}_{\omega_j\beta m}(\rho_j) %
+d_{jm}\,\tilde{\mathbf{N}}_{\omega_j\beta m}(\rho_j)\Bigr), %
\qquad\rho_j<R.
\end{array}
\label{AppS_EHint}%
\end{equation}
Here $\omega_j=n_j\omega$, and $\rho_j$, $\phi_j$ are the
cylindrical coordinates of the vector $\mathbf{r}-\mathbf{r}_j$,
where $\mathbf{r}_j$ is the coordinates of the axis of the $j$-th
waveguide. The vector cylinder harmonics
$\tilde{\mathbf{M}}_{\omega_j\beta m}(\rho_j)$ and
$\tilde{\mathbf{N}}_{\omega_j\beta m}(\rho_j)$ are defined as
follows
\begin{equation}
\tilde{\mathbf{N}}_{\omega_j\beta m}(\rho_j) %
=\mathbf{e}_r\,\frac{i\beta}{\varkappa_j}\,J'_m(\varkappa_j\rho_j) %
-\mathbf{e}_\phi\,\frac{m\beta}{\varkappa_j^2\rho_j}\,J_m(\varkappa_j\rho_j) %
+\mathbf{e}_z\,J_m(\varkappa_j\rho_j), %
\label{tildeN}
\end{equation}
\begin{equation}
\tilde{\mathbf{M}}_{\omega_j\beta m}(\rho_j) %
=\mathbf{e}_r\,\frac{m\omega_j}{\varkappa_j^2\rho_j}\,J_m(\varkappa_j\rho_j) %
+\mathbf{e}_\phi\,\frac{i\omega_j}{\varkappa_j}\,J'_m(\varkappa_j\rho_j), %
\label{tildeM}
\end{equation}
where $\varkappa_j=\sqrt{\omega_j^2-\beta^2}$,
$J_m(\varkappa_j\rho_j)$ is the Bessel function, and the prime
means the derivative with respect to the argument
$\varkappa_j\rho_j$.

Let us turn to the electromagnetic field outside of the array.
This field is the sum of the contributions of all the waveguides,
\begin{equation}
\mathbf{E}(\mathbf{r})=\sum\limits_{j=1}^N \mathbf{E}_j(\mathbf{r}), %
\qquad %
\mathbf{H}(\mathbf{r})=\sum\limits_{j=1}^N \mathbf{H}_j(\mathbf{r}). %
\label{Sum11}
\end{equation}
The contribution induced by the $j$-th waveguide and vanishing at
$\rho_j\to\infty$ may be represented in the form
\begin{equation}
\begin{array}[c]{l} \displaystyle
\mathbf{E}_j(\mathbf{r})= %
e^{-i\omega t+i\beta z}\sum\limits_m e^{im\phi_j}\, %
\Bigl(a_{jm}\,\mathbf{N}_{\omega_e\beta m}(\rho_j) %
-b_{jm}\,\mathbf{M}_{\omega_e\beta m}(\rho_j)\Bigr),
\medskip \\ \displaystyle
\mathbf{H}_j(\mathbf{r})= %
e^{-i\omega t+i\beta z} n_e \sum\limits_m e^{im\phi_j}\, %
\Bigl(a_{jm}\,\mathbf{M}_{\omega_e\beta m}(\rho_j) %
+b_{jm}\,\mathbf{N}_{\omega_e\beta m}(\rho_j)\Bigr), %
\qquad\rho_j>R.
\end{array}
\label{EX_J}
\end{equation}
Here $\omega_e=n_e\omega$. The vector cylinder harmonics
$\tilde{\mathbf{M}}_{\omega_j\beta m}(\rho_j)$ and
$\tilde{\mathbf{N}}_{\omega_j\beta m}(\rho_j)$ are
\begin{equation}
\mathbf{N}_{\omega_e\beta m}(\rho_j)= %
\mathbf{e}_r\,\frac{i\beta}{\varkappa_e}\,H'_m(\varkappa_e\rho_j) %
-\mathbf{e}_\phi\,\frac{m\beta}{\varkappa_e^2\rho_j}\,H_m(\varkappa_e\rho_j) %
+\mathbf{e}_z\,H_m(\varkappa_e\rho_j), %
\label{N}
\end{equation}
\begin{equation}
\mathbf{M}_{\omega_e\beta m}(\rho_j)= %
\mathbf{e}_r\,\frac{m\omega_e}{\varkappa_e^2\rho_j}\,H_m(\varkappa_e\rho_j) %
+\mathbf{e}_\phi\,\frac{i\omega_e}{\varkappa_e}\,H'_m(\varkappa_e\rho_j), %
\label{M}
\end{equation}
where $\varkappa_e=\sqrt{\omega_e^2-\beta^2}$, and
$H_m(\varkappa_e\rho_j)$ is the Hankel function of the first kind.
Note that, for $\beta=0$, Eqs. (\ref{AppS_EHint}) and (\ref{EX_J})
transform into the corresponding expressions in \cite{VanDeHulst},
however different notations are used there.


Below we consider the simplest approximation to these equations,
namely, the zero-harmonic approximation. This means that in Eqs.
(\ref{AppS_EHint}) and (\ref{EX_J}) only the terms with $m=0$ are
taken into account. In this approximation there are two kinds of
the guided modes, namely the transverse magnetic (TM) and
transverse electric (TE) modes. For the TM-mode $b_{j0}=d_{j0}=0$,
and for the TE-mode $a_{j0}=c_{j0}=0$. As an example, let us
consider the TM-modes.
\begin{equation}
\tilde{\mathbf{E}}_j(\mathbf{r})= %
c_j\,\tilde{\mathbf{N}}_{\omega_j\beta 0}(\rho_j), %
\qquad %
\tilde{\mathbf{H}}_j(\mathbf{r})= %
c_j\,\tilde{\mathbf{M}}_{\omega_j\beta 0}(\rho_j) %
\qquad\rho_j<R. %
\label{EHint_ZHA}
\end{equation}
\begin{equation}
\mathbf{E}_j(\mathbf{r})= %
a_j\,\mathbf{N}_{\omega_e\beta 0}(\rho_j), %
\qquad %
\mathbf{H}_j(\mathbf{r})= %
a_j\,\mathbf{M}_{\omega_e\beta 0}(\rho_j) %
\qquad\rho_j>R. %
\label{EHex_ZHA}
\end{equation}
Here and below, $a_j$ and $c_j$ stand for $a_{j0}$ and $c_{j0}$,
and the factor $e^{-i\omega t+i\beta z}$ is omitted.

To derive the equations that determine the partial amplitudes
$a_j$ and $c_j$, one should use the boundary conditions on the
surface of every waveguide. The boundary conditions on the surface
of the $j$-th waveguide connect the field %
$\tilde{\mathbf{E}}_j(\mathbf{r})$, %
$\tilde{\mathbf{H}}_j(\mathbf{r})$ %
inside the $j$-th waveguide and the field %
$\mathbf{E}(\mathbf{r})$, $\mathbf{H}(\mathbf{r})$ %
outside. In general, there are four independent boundary
conditions. However, for $m=0$ TM-modes and only two boundary
conditions are required:
\begin{equation}
\left[\mathbf{E}(\mathbf{R}_j)\right]_z= %
\left[\tilde{\mathbf{E}}_j(\mathbf{R}_j)\right]_z, %
\qquad %
\left[\mathbf{H}(\mathbf{R}_j)\right]_\phi= %
\left[\tilde{\mathbf{H}}_j(\mathbf{R}_j)\right]_\phi, %
\label{MSF_BoundCond}
\end{equation}
here $\mathbf{R}_j$ is the radius-vector of a point on the surface
of the $j$-th waveguide.

The boundary conditions on the surface of the $j$-th waveguide can
be expressed in the most convenient form if the contributions of
all the waveguides to the field outside the array are expressed in
terms of the same argument $\rho_j$. For this propose we apply the
following relations:
\begin{equation}
\begin{array}{l} \displaystyle
\mathbf{N}_{\omega_e\beta 0}(\rho_l)\approx %
U_{lj}(\omega,\beta)\, %
\tilde{\mathbf{N}}_{\omega_e\beta 0}(\rho_j),
\medskip \\ \displaystyle
\mathbf{M}_{\omega_e\beta 0}(\rho_l)\approx %
U_{lj}(\omega,\beta)\, %
\tilde{\mathbf{M}}_{\omega_e\beta 0}(\rho_j),
\qquad l\neq j,
\end{array}
\label{MSF_Trans}
\end{equation}
where %
$U_{lj}(\omega,\beta)=H_0(\varkappa_e r_{lj})$, %
$H_0$ being the Hankel function of the first kind and $r_{lj}$
being the distance between the axes of the $j$-th and the $l$-th
waveguides. The relations (\ref{MSF_Trans}) follow from the Graf
theorem (see \cite{AbramowitzStegun}) in the zero-harmonic
approximation.

Thus, it follows from (\ref{EHex_ZHA}) that
\begin{equation}
\mathbf{E}_l(\mathbf{r})= %
a_j\,U_{lj}(\omega,\beta)\,\tilde{\mathbf{N}}_{\omega_e\beta 0}(\rho_j), %
\qquad %
\mathbf{H}_l(\mathbf{r})= %
a_j\,n_e\,U_{lj}(\omega,\beta)\,\tilde{\mathbf{M}}_{\omega_e\beta 0}(\rho_j). %
\label{EHex_ZHA01}
\end{equation}
Substituting (\ref{EHex_ZHA01}) to (\ref{Sum11}), one gets
\begin{equation}
\begin{array}{l} \displaystyle
\mathbf{E}(\mathbf{r})= %
a_j\,\mathbf{N}_{\omega_e\beta 0}(\rho_j) %
+\sum\limits_{l\neq j}
a_l\,U_{lj}(\omega,\beta)\, %
\tilde{\mathbf{N}}_{\omega_e\beta 0}(\rho_j),
\medskip \\ \displaystyle
\mathbf{H}(\mathbf{r})= %
a_j\,n_e\,\mathbf{M}_{\omega_e\beta 0}(\rho_j) %
+\sum\limits_{l\neq j}
a_l\,n_e\,U_{lj}(\omega,\beta)\, %
\tilde{\mathbf{M}}_{\omega_e\beta 0}(\rho_j).
\end{array}
\label{Sum11_01}
\end{equation}

So, the boundary conditions (\ref{MSF_BoundCond}) take the form
\begin{equation}
\begin{array}{l} \displaystyle
a_j\,H_0(\varkappa_e R) %
+\sum\limits_{l\neq j} %
a_l\,U_{lj}(\omega,\beta)\,J_0(\varkappa_e R)= %
c_j\,J_0(\varkappa_j R),
\medskip \\ \displaystyle
a_j\,i\frac{n_e\omega_e}{\varkappa_e}\,H_0'(\varkappa_e R) %
+\sum\limits_{l\neq j} %
a_l\,U_{lj}(\omega,\beta)\,i\frac{n_e\omega_e}{\varkappa_e}\,J_0'(\varkappa_e R)= %
c_j\,i\frac{n_j\omega_j}{\varkappa_j}\,J_0'(\varkappa_j R).
\end{array}
\label{MSF_Syst01}
\end{equation}

Eqs. (\ref{MSF_Syst01}) lead to the following system of equations:
\begin{equation}
\frac{a_j}{\bar{a}_j(\omega,\beta)} %
-\sum\limits_{l\neq j} %
U_{jl}(\omega,\beta)\,a_l=0, %
\label{MSF_MainSyst}
\end{equation}
\begin{equation}
c_j=\bar{c}_j(\omega,\beta)\,a_j. %
\label{MSF_cj0_aj0}
\end{equation}
Here
\begin{equation}
\bar{a}_j(\omega,\beta)= %
\frac{n_j^2 \varkappa_e\,J_0'(\varkappa_j R)\,J_0(\varkappa_e R) %
-n_e^2\varkappa_j\,J_0(\varkappa_j R)\,J_0'(\varkappa_e R)} %
{n_e^2\varkappa_j\,J_0(\varkappa_j R)\,H_0'(\varkappa_e R) %
-n_j^2 \varkappa_e\,J_0'(\varkappa_j R)\,H_0(\varkappa_e R)}, %
\label{MSF_a}
\end{equation}
\begin{equation}
\bar{c}_j(\omega,\beta)= %
\frac{n_e^2\varkappa_j\{H_0(\varkappa_e R)\,J_0'(\varkappa_e R) %
-H_0'(\varkappa_e R)\,J_0(\varkappa_e R)\}} %
{n_e^2\varkappa_j\,J_0'(\varkappa_e R)\,J_0(\varkappa_j R) %
-n_j^2\varkappa_e\,J_0(\varkappa_e R)\,J_0'(\varkappa_j R)}. %
\label{MSF_c}
\end{equation}


\section{Method for the optical beam calculation.}
\label{Sec:CalcMethod}

The system of equations (\ref{MSF_MainSyst}) describes the guided
modes of the array of waveguides. This system possesses the
nontrivial solution only if the determinant of the matrix of this
system vanishes,
\begin{equation}
\det\left\|\frac{\delta_{jl}}{\bar{a}_j(\omega,\beta)} %
-U_{jl}(\omega,\beta)\right\|=0. %
\label{MSF_det}
\end{equation}
This equation allows to obtain the propagation constants $\beta_n$
of the guided modes for the given frequency $\omega$, $n$ being
the number of a guided mode. There are $N$ solutions of Eq.
(\ref{MSF_det}).

Let $a_j(\beta_n)$ be the normalized solution of Eq. (\ref{MSF_MainSyst}), %
$\sum\limits_{j=1}^N\vert a_j(\beta_n)\vert^2=1$. %
The guided mode of the frequency $\omega$ is a superposition of
the modes with the different $\beta_n$:
\begin{equation}
\begin{array}{l} \displaystyle
\mathbf{E}(t,\mathbf{r})=e^{-i\omega t}\, %
\sum\limits_n C_n\,e^{i\beta_n z} %
\sum\limits_{j=1}^N a_j(\beta_n)\, %
\mathbf{N}_{\omega_e\beta_n 0}(\rho_j), %
\medskip \\ \displaystyle
\mathbf{H}(t,\mathbf{r})=n_e\,e^{-i\omega t}\, %
\sum\limits_n C_n\,e^{i\beta_n z} %
\sum_{j=1}^N a_j(\beta_n)\, %
\mathbf{M}_{\omega_e\beta_n 0}(\rho_j). %
\end{array}
\label{MSF_LinearSuperposition}
\end{equation}
The coefficients $C_n$ determine the superposition.

The functions %
$\mathbf{N}_{\omega_e\beta_n 0}(\rho_j)$, %
$\mathbf{M}_{\omega_e\beta_n 0}(\rho_j)$ %
vanish rapidly as $\rho_j$ increases. So, the field near the
$j$-th waveguide is mainly determined by the partial amplitudes
$a_j(\beta_n)$. Thus, for the field near the $j$-th waveguide one
can retain in (\ref{MSF_LinearSuperposition}) only the terms which
contain $a_j(\beta_n)$:
\begin{equation}
\begin{array}{l} \displaystyle
\mathbf{E}(t,\mathbf{r})=e^{-i\omega t}\, %
\sum\limits_n C_n\,e^{i\beta_n z}\, %
a_j(\beta_n)\,\mathbf{N}_{\omega_e\beta_n 0}(\rho_j), %
\medskip \\ \displaystyle
\mathbf{H}(t,\mathbf{r})=n_e\,e^{-i\omega t}\, %
\sum\limits_n C_n\,e^{i\beta_n z}\, %
a_j(\beta_n)\,\mathbf{M}_{\omega_e\beta_n 0}(\rho_j). %
\end{array}
\label{MSF_LinearSuperposition01}
\end{equation}
If the waveguides interact weakly ($U_{lj}(\omega,\beta)\ll
1/\bar{a}_j(\omega,\beta)$) and the difference between the
waveguides is negligible ($n_j-n_{j-1} \ll n_j$), all the values
$\beta_n$ are close to each other. For this case,
$\mathbf{N}_{\omega_e\beta_n 0}(\rho_j)$ and
$\mathbf{M}_{\omega_e\beta_n 0}(\rho_j)$ in
(\ref{MSF_LinearSuperposition01}) may be approximately replaced
with $\mathbf{N}_{\omega_e \bar{\beta} 0}(\rho_j)$ and
$\mathbf{M}_{\omega_e \bar{\beta} 0}(\rho_j)$, where
$\bar{\beta}=\left(\sum\limits_n \beta_n\right)/N$. Thus, instead
of (\ref{MSF_LinearSuperposition01}) one obtains
\begin{equation}
\begin{array}{l} \displaystyle
\mathbf{E}(t,\mathbf{r})=e^{-i\omega t}\, %
\sum\limits_n C_n\,e^{i\beta_n z}\, %
a_j(\beta_n)\,\mathbf{N}_{\omega_e \bar{\beta} 0}(\rho_j)= %
e^{-i\omega t}\,A_j(z)\,\mathbf{N}_{\omega_e \bar{\beta} 0}(\rho_j), %
\medskip \\ \displaystyle
\mathbf{H}(t,\mathbf{r})=n_e\,e^{-i\omega t}\, %
\sum\limits_n C_n\,e^{i\beta_n z}\, %
a_j(\beta_n)\,\mathbf{M}_{\omega_e \bar{\beta} 0}(\rho_j)= %
n_e\,e^{-i\omega t}\,A_j(z)\,\mathbf{M}_{\omega_e \bar{\beta} 0}(\rho_j). %
\end{array}
\label{MSF_LinearSuperposition02}
\end{equation}
In (\ref{MSF_LinearSuperposition02}) the modal amplitudes $A_j(z)$
are introduced,
\begin{equation}
A_j(z) =\sum\limits_n C_n\,e^{i\beta_n z}\,a_j(\beta_n). %
\label{iii}
\end{equation}
For the case of weakly interacting nearly identical waveguides,
the modal amplitudes $A_j(z)$ represent the behavior of the guided
modes properly. Below we define the optical excitation intensity
at the point with the coordinate $z$ of the $j$-th waveguide as
$|A_j(z)|$.

The coefficients $C_n$ are obtained from the boundary condition at
$z=0$:
\begin{equation}
\sum\limits_n\,C_n\,a_j(\beta_n)=A_j(0). %
\label{a4}
\end{equation}
The system of equations (\ref{a4}) allows to obtain the
coefficients $C_n$ for given $A_j(0)$.

The boundary condition $A_j(0)$ are determined with the input wave
packet which possesses the Gaussian form. Thus,
\begin{equation}
A_j(0)=e^{-\frac{j^2}{\sigma^2}+ik_0 aj}. %
\label{a5}
\end{equation}
This means that the input wave packet approximately illuminates
the ends of the waveguides with the numbers $-\sigma<j<\sigma$,
and the with of the input wave packet is $W=2\sigma a$. The phase
difference between the amplitudes taken at the ends of the nearest
waveguides is $k_0 a$.

Thus, the guided mode can be found as follows:

1) Calculate numerically the set of propagating constants
$\beta_n$ using Eq. (\ref{MSF_det});

2) Obtain the amplitudes $a_j(\beta_n)$ for every $\beta_n$ using
Eq. (\ref{MSF_MainSyst});

3) For the given boundary conditions find the coefficients $C_n$
using Eq. (\ref{a4});

4) Calculate the function $A_j(z)$ by means of Eq. (\ref{iii}).


\section{The optical beam in the array of waveguides for different boundary conditions.}
\label{Sec:NumericalSimulation}

We apply the developed technique for calculating the optical beam
in an array represented in Fig. \ref{Fig01_Scheme}. We consider a
sample that can be fabricated by means of the technology
represented in \cite{Withford2013, Withford2013a}. In these works
a new method to fabricate low bend loss femtosecond-laser written
waveguides is explained. The parameters taken for the numerical
simulation correspond approximately to the parameters of the
arrays of waveguides reported in work \cite{Withford2013}.

\begin{figure}
\centering
\includegraphics[width=0.7\textwidth]{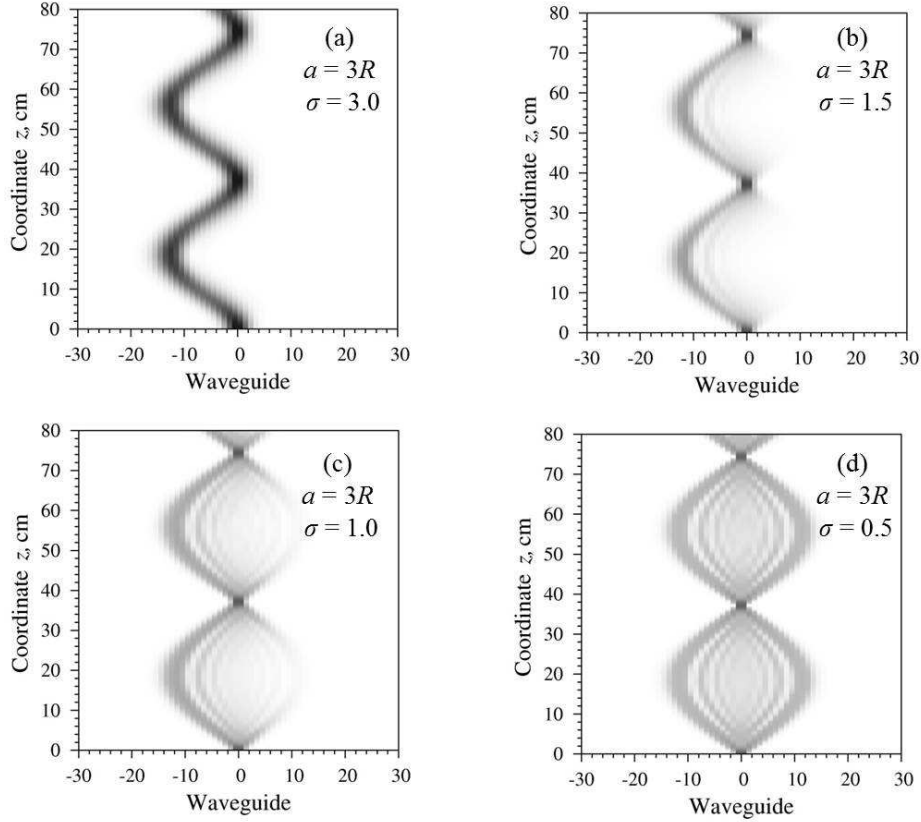} %
\caption{The optical beam in the array for different boundary conditions. %
The period of the array $a=3R$. The width of the input wave
packet: (a) $\sigma=3.0$, (b) $\sigma=1.5$, (c) $\sigma=1.0$,
(d) $\sigma=0.5$.} %
\label{Fig02_Result_30R}
\end{figure}

\begin{figure}
\centering
\includegraphics[width=0.7\textwidth]{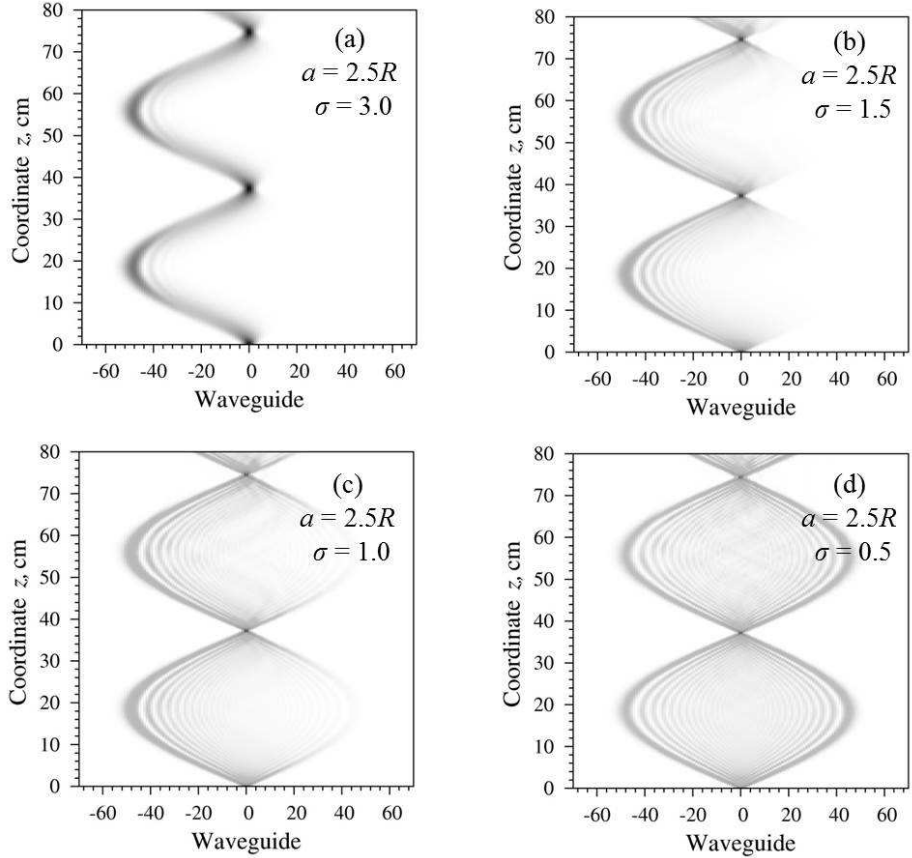} %
\caption{The optical beam in the array for different boundary conditions. %
The period of the array $a=2.5R$. The width of the input wave
packet: (a) $\sigma=3.0$, (b) $\sigma=1.5$, (c) $\sigma=1.0$,
(d) $\sigma=0.5$.} %
\label{Fig03_Result_25R}
\end{figure}

The wavelength of the laser source in \cite{Withford2013} is
$\lambda=1550~\mathrm{nm}$. We take the waveguide radius
$R=5\lambda=7750~\mathrm{nm}$. The refractive index of the
environment is $n_e=1.4877$, and the refractive index of the
waveguide $j=0$ (the central waveguide of the array) is %
$n_0=n_e+5\times 10^{-3}$. %
These parameters also approximately correspond to the experiments
reported in \cite{Withford2013}. For our calculations we assume
that the variation of refractive indices between the nearest
waveguides is $\delta n=n_j-n_{j-1}=5\times 10^{-6}$. We take the
arrays with two different periods $a_1=3R$ and $a_2=2.5R$.

We produce the calculation for different values of the input wave
packet width: $\sigma_1=3.0$, $\sigma_2=1.5$, $\sigma_3=1.0$,
$\sigma_4=0.5$. In all cases $k_0=0$.

The results of the numerical simulation for $a_1=3R$ and
$a_2=2.5R$ are presented in Fig. \ref{Fig02_Result_30R} and Fig.
\ref{Fig03_Result_25R} correspondingly.

The figures show that generally similar qualitative results are
obtained for the both selected periods $a$ of the array. Namely,
$\sigma=3$ (i. e. the input wave packet illuminates nearly seven
waveguides) is enough wide for the optical Bloch oscillations
occur. At the same time, the optical beam obtained for
$\sigma=0.5$ doubtless can be considered as the breathing mode.

But for $\sigma=1.5$ and $\sigma=1.0$ the form of the optical beam
does not relate any of these phenomena. The most exactly, the
obtained picture can be defined as an ``asymmetrical breathing
modes''. In fact, the form of the optical beam resembles the
breathing mode, but the left side of the beam is brighter then the
right side, and the intensity of the beam gradually decreases with
the shift to the right. When $\sigma$ increases, the intensity
reduction in the transverse direction becomes faster. Thus, as
$\sigma$ becomes large enough, the optical beam concentrates along
the oscillating curve, which for small $\sigma$ formes the left
edge of the breathing mode. Thus, the ``asymmetrical breathing
mode'' turns into the optical Bloch oscillations.


\section{Conclusion}
\label{Sec:Conclusion}

In this paper we investigated the optical beam behavior in the
array of the coupled optical waveguides with the monotonic change
of the refractive index in the transverse direction. We considered
the optical beam form dependence on the boundary conditions, i.e.
on the form of the input wave packet.

For this purpose we applied the multiple scattering formalism
based on the macroscopic electrodynamics approach. For the
simplicity, we used the zero-harmonic approximation taking into
account the TM-modes only. The MSF is the basis of the numerical
algorithm for calculating the form of the optical beam for the
specified boundary conditions at $z=0$.

We chose the boundary conditions possessing the form of the
Gaussian wave packet. Varying the width of the input wave packet,
we observed the change of the optical beam excited in the array.

As expected, for the input wave packet being narrow enough, the
breathing mode was observed, while for the case of enough large
input wave packet, the optical Bloch oscillation occurred. But for
the intermediate case, the optical beam takes an unexpected form
which can be considered neither as the Bloch oscillations nor as
the breathing mode. The obtained form of the optical beam can be
defined as an ``asymmetrical breathing mode''.  We qualitatively
described the optical beam transformation when the width of the
input wave packet changes. We showed that the breathing mode and
the optical Bloch oscillations are the limiting cases of the
``asymmetrical breathing mode''.


\subsection*{Acknowledgments}

The study is supported by the Russian Fund for Basic Research
(Grant 14-29-08165).


\end{document}